\documentclass[pre,twocolumn,floatfix,superscriptaddress,footinbib, sort&compress, numbers, merge, reprint]{revtex4-1}
\usepackage{graphicx}
\usepackage{amsmath}
\usepackage{amssymb}
\usepackage{booktabs}
\usepackage{color}
\usepackage{natbib}

\usepackage{chngcntr}

\usepackage[utf8]{inputenc}
\usepackage[tight,sf]{subfigure}%

\newcommand{\be}{\begin{eqnarray}}
\newcommand{\ee}{\end{eqnarray}}

\begin{document}

\title{Controlling fracture cascades through twisting and quenching}

\author{Ronald H. Heisser}
\thanks{Current address: Sibley School of Mechanical and Aerospace Engineering
Cornell University, 
105 Upson Hall, 
Ithaca, New York 14853}
\affiliation{Department of Mechanical Engineering, 
Massachusetts Institute of Technology, 77 Massachusetts Avenue, Cambridge,~MA~02139, USA}

\author{Vishal P. Patil}
\thanks{Joint first author}
\affiliation{Department of Mathematics, Massachusetts Institute of Technology, 77 Massachusetts Avenue, Cambridge,~MA~02139, USA}

\author{Norbert Stoop}
\affiliation{Department of Mathematics, Massachusetts Institute of Technology, 77 Massachusetts Avenue, Cambridge,~MA~02139, USA}

\author{J\"orn Dunkel} 
\affiliation{Department of Mathematics, Massachusetts Institute of Technology, 77 Massachusetts Avenue, Cambridge,~MA~02139, USA}
\date{\today}

\pacs{}

\begin{abstract}
Fracture limits the structural stability of macroscopic and microscopic materials, from
beams and bones to microtubules and nanotubes.  Despite recent progress, fracture control continues to present profound practical and theoretical challenges. A famous longstanding problem posed by Feynman asks why brittle elastic rods appear almost always to fragment into at least three pieces when placed under large bending stresses. Feynman's observation raises fundamental questions about the existence of protocols that can robustly induce binary fracture in brittle materials. Using experiments, simulations and analytical scaling arguments, we demonstrate controlled binary fracture of brittle elastic rods for two distinct protocols based on twisting and nonadiabatic quenching. Our experimental data for twist-controlled fracture agree quantitatively with a theoretically predicted phase diagram. Furthermore, we establish novel asymptotic scaling relations for quenched fracture. Due to their generality, these results are expected to apply to torsional and kinetic fracture processes in a wide range of systems.
\end{abstract}

\maketitle

\section{Introduction}

Elastic rods (ERs) are ubiquitous in natural and man-made matter, performing important physical and biological functions across a wide range of scales, from  columns~\cite{Beams_Book}, trees~\cite{Virot2016} and bones~\cite{Peterlik:2006aa} to the legs of water striders~\cite{Hu:2003aa}, semi-flexible polymer~\cite{vasily_stretch_coil} networks~\cite{Sanchez12,2014Broedersz_RMP}  and carbon nanotube composites~\cite{Baughman787}. When placed under extreme stresses, the structural stability of such materials becomes ultimately limited by the fracture behaviors of their individual fibrous or tubular  constituents. Owing to their central practical importance in engineering, ER fracture and crack propagation have been intensively studied for more than a century both experimentally~\cite{1997Bouchaud,Audoly05,Gladden05} and theoretically~\cite{1988Herrmann,1998Peerlings,2007GerstleSilling}. Recent advances in video microscopy  and microscale force manipulation~\cite{2001Evans,Neuman:2008aa} have extended the scope of fracture studies to the microworld~\cite{1997Gilbert_APL,2011Demetriu_Glass}, revealing causes and effects of structural failure in  the axonal cytoskeleton~\cite{Tang-Schomer10}, fibroblasts~\cite{Odde99}, bacterial flagellar motors~\cite{2008Attmannspacher}, active liquid crystals~\cite{Sanchez12} and multi-walled carbon nanotubes~\cite{2000Yu_Science, Mokashi07}. 
\par
Although important theoretical progress~\cite{Audoly05,Gladden05,Wittel04,Mitchell17} has been achieved over the past two decades, even basic qualitative aspects of the fracture phenomenology remain poorly understood. Bending induced ER fragmentation has been thoroughly investigated in the limits of adiabatically slow~\cite{Audoly05} and diabatically fast~\cite{Gladden05} energy injection, but the roles of twist and quench rate on the fracture process have yet to be clarified. These two fundamental issues are directly linked to a famous observation by Richard Feynman~\cite{Sykes96}, who noted that dry spaghetti, when brought to fracture by holding the ends and moving them towards each other, appears almost always to break into at least three pieces. The phenomenon of non-binary ER fracture is also well known to pole vaulters, with a notable instance occuring during the 2012 Olympic Games~\cite{Olympics}. Below, we will revisit and generalize Feynman's experiment, in order to investigate  systematically how twist and quench dynamics influence the elastic  fragmentation cascade\cite{Audoly05,Gladden05}. Specifically, we will demonstrate two complementary quench protocols for controlled binary fracture of brittle ERs, thereby identifying conditions  under which Feynman's fragmentation conjecture becomes invalid. Our experimental observations are in good agreement with numerical predictions from a nonlinear elasticity model, and can be rationalized through analytical scaling arguments.

\begin{figure*}[t!]
\includegraphics[width=2\columnwidth]{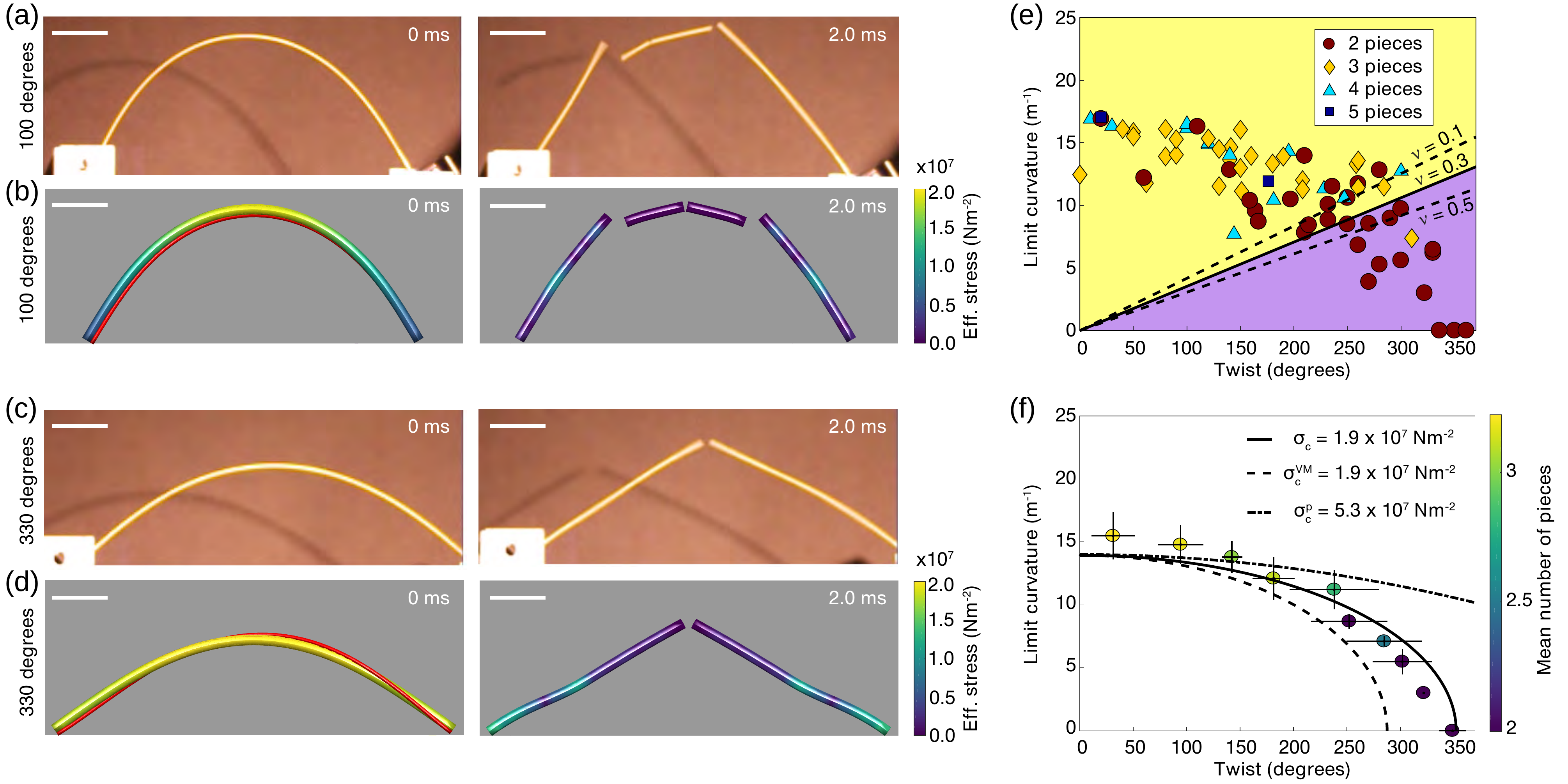}
\caption{\textbf{Using twist to break Feynman's fragmentation bound.}
(a)~High-speed images from an experiment with subcritical twist angle showing fragmentation into more than two pieces, in agreement with Feynman's conjecture~(Movie 1). Time $t=0$ (left) is defined as the moment (last frame) before fracture
(b)~Simulations also predict fracture in more than two pieces for parameters corresponding to the experiment in~(a). Due to perfectly symmetric initial conditions, our simulations generally produce even fragment numbers (Movie 1). Red line illustrates twist. 
(c)~At supercritical twist angles, the maximum curvature before fracture is significantly lowered enabling twist-controlled binary fracture (Movie 2). 
(d)~Simulations for the experimental parameters in (c) also confirm binary fracture (Movie 2). 
(e)~Phase diagram showing  that binary fracture dominates for twist angles larger than $\sim 250$ degrees (73 data points in total). The theoretically predicted region (purple) in which an ideal rod is expected to exhibit binary fracture depends only weakly on the Poisson's ratio~$\nu$ and agrees well with  the data.
(f)~Experimental data from (e), averaged over 10 sectors defined by the rays $(\mu^2JL^{-1}\cos\left(n\pi/2\right),E^2I\sin\left(n\pi/2\right))$ for $n=0,1\dots 11$, follow the theoretically predicted critical ellipse (solid curve) from Eq.~\eqref{twistyield}. The dashed curve shows the von Mises ellipse from Eq.~\eqref{vonMises}, and the dash-dotted curve shows the parabola of constant maximum principal stress from Eq.~\eqref{eigenStress}. The data in Fig. \ref{speed} yield $\sigma_c=\sigma_c^\mathrm{VM}=1.9 \times 10^{7}\,\text{N/m}^2$ and $\sigma^P_c = 5.3\times 10^7\,\text{N/m}^2$ at zero twist.  Error bars show standard deviations.
The critical curvatures for simulations at different twist  in (b) and (d) are chosen according to the critical stress ellipse in (f), and the minimum fragment length  $\lambda=3\,$cm  estimated from the data in Fig. \ref{speed}. Diameter of rod in (b,d) enhanced for visualization. Scale bars in (a-d) $3\,$cm. 
}
\label{twist}
\end{figure*}


\section{Results}
\textbf{Model.}
We describe an ER at time $t$ by its arc length parametrized centerline $\mathbf{x}(s,t)$, $s\in [0,L]$, and an orthonormal frame $\{\mathbf{d_1}(s,t),\mathbf{d_2}(s,t),\mathbf{d_3}(s,t)\}$ such that $\mathbf{d_3}=\mathbf{x}'$, where primes denote $s$-derivatives and dots denote $t$-derivatives. We further assume the rod is uniform with density $\rho$, naturally straight and inextensible with circular cross sectional area $A=\pi r^2$. Its moment of inertia $I$ and moment of twist $J$ are then given by\cite{Audoly10} $J=2I=\pi r^4/2$. The rod's dynamics are governed by the damped Kirchhoff equations \cite{Coleman93, Audoly10} (SI)
\begin{subequations}\label{Kirch}
\begin{eqnarray}
\mathbf{F''} &=& \rho A\ddot{\mathbf{d}}_3 \\
\mathbf{M}' + \mathbf{d_3\times F} &=& \dot{\mathbf{L}}+4b\rho I\omega_3\mathbf{d_3} 
\end{eqnarray}
\end{subequations}
where $\mathbf{F}(s,t)$ is the force, $\mathbf{M}(s,t)=EI\kappa_1\mathbf{d_1}+EI\kappa_2\mathbf{d_2}+\mu J\kappa_3\mathbf{d_3}$ is the internal moment, $\mathbf{L}(s,t) = \rho I \omega_1\mathbf{d_1}+\rho I\omega_2\mathbf{d_2}+2\rho I\omega_3\mathbf{d_3}$ is the cross sectional angular momentum, and the vectors $\boldsymbol{\kappa}=\kappa_i\mathbf{d_i},\; \boldsymbol{\omega}=\omega_i\mathbf{d_i}$ satisfy $\mathbf{d_i}' = \boldsymbol{\kappa}\times \mathbf{d_i},\; \dot{\mathbf{d_i}} = \boldsymbol{\omega}\times\mathbf{d_i}$.  The Young's modulus $E$ and the shear modulus $\mu$ are related by $E/\mu = 2(1+\nu)$, where $\nu$ is the Poisson's ratio. Most materials have $0.2<\nu<0.5$. The final term on the rhs. of Eq.~\ref{Kirch}(b) denotes damping of twist modes with damping parameter $b$. Our measurements of this parameter using a torsion pendulum (Appendix) indicate that twist is approximately critically damped (SI). Since the timescale for the entire fracture cascade is an order of magnitude smaller than the time period of the fundamental bending mode, we do not need to include bending damping terms in our analysis. The average material properties of our experimental samples are $2r=1.4\pm 0.1\,$mm, $\rho= 1.5\pm 0.1\,\textrm{g/cm}^{3}$, $E=3.8\pm 0.3\,$GPa, $\mu=1.5\pm 0.2\,$GPa, and by considering mean values for $E,\mu$ we obtain $\nu=0.3\pm 0.1$ (Appendix). Finally, we note that the Kirkhhoff equations do not account for certain shear effects described by Timoshenko beam theory. Indeed, the Timoshenko theory does provide a more accurate description of bending waves with large wavenumber compared to rod radius. However, to describe fracture, we will only need to consider wavenumbers $k$ with $kr/2\pi < 0.1$. In this regime, the difference between the Timoshenko and Kirkhhoff beam theories is negligible \cite{Waves_Book}.
\par
To compare individual experiments with theoretical predictions, we solve the Kirchhoff equations~\eqref{Kirch} numerically with a discrete differential geometry algorithm~\cite{Bergou08, Bergou10} (Appendix), adopting a stress-based fracture criterion defined as follows: We define the twist of the rod, $\theta(s,t)$, from the twist density, $\kappa_3(s,t)$, by $\theta'=\kappa_3$. The effective stress at a point, $\sigma(s)$, is obtained by integrating a scalar invariant of the full stress tensor,~$\boldsymbol{\hat{S}}$, over a cross section of the rod
\begin{equation}\label{stress}
\begin{aligned}
\sigma(s,t)^2 &= \frac{1}{2\pi r^2}\int\mathrm{tr}\left( \boldsymbol{\hat{S}}^\top\boldsymbol{\hat{S}}\right)dA
 \\
&= \frac{1}{4}E^2 r^2 \kappa(s,t)^2 + \frac{1}{2}\mu^2 r^2 \theta'(s,t)^2 
\end{aligned}
\end{equation}
where $\kappa = \left(\kappa_1^2+\kappa_2^2\right)^{1/2}$ is the geometrical curvature of the centerline and $\mathrm{tr}$ denotes the trace.
If the rod is in a steady state, the twist density is constant, $\theta'=Tw/L$, where $Tw$ is the total applied twist.
We posit that the rod fractures at a point $s$ along the curve if the effective stress $\sigma(s)$ exceeds a critical value $\sigma_c$, and that no two fractures can occur within a minimal fracture distance\cite{Grady10} $\lambda$ of each other.  For a uniform twist distribution, the critical stress imposes a critical yield curvature~$\kappa_c$ by
\begin{eqnarray}\label{twistyield}
\sigma_c^2=\frac{1}{4}E^2r^2\kappa_c^2 + \frac{1}{2}\mu^2 r^2 \left(\frac{Tw}{L}\right)^2.
\end{eqnarray} 
For comparison, by integrating the classical von Mises stress criterion over a cross section, we obtain a critical local stress ellipse given by (SI) 
\begin{eqnarray}\label{vonMises}
\left(\sigma^{\mathrm{VM}}_c\right)^2= \frac{1}{4}E^2r^2 \kappa_c^2 + \frac{3}{4}\mu^2 r^2 \left(\frac{Tw}{L}\right)^2
\end{eqnarray}
Another common criterion comes from considering the maximum eigenvalue of the stress tensor, or maximum principal stress, on the boundary of the rod. This gives a critical stress parabola
\begin{multline}\label{eigenStress}
\left(\sigma_c^{p}\right)^2 = \mu^2r^2\left(\frac{Tw}{L}\right)^2 + E^2r^2\kappa_c^2 +\\ +Er^2\kappa_c \sqrt{E^2\kappa_c^2+2\mu^2\left(\frac{Tw}{L}\right)^2}
\end{multline}
All three curves are qualitatively consistent with our data, with Eq.~\eqref{twistyield} 
yielding the best quantitative agreement (Fig.~\ref{twist}f).
\par
Our model contains exactly one free parameter, $\lambda$, which we introduce to account for the fact that the Kirchhoff equations become invalid over the small length scales and time scales near the fracture tip. As shown by Audoly and Neukirch\cite{Audoly05}, when an initially uniformly curved ER is released from one end, its local curvature increases at the free end. When a rod fractures at a point of maximum curvature, the Kirchhoff model possesses solutions in which the curvature near the fracture tip increases even further; in the case of  $\lambda=0$, this would trigger additional fractures arbitrarily close to the first fracture, which is not observed experimentally.  Following standard fragmentation theory\cite{Grady10, Villermaux07}, we therefore assume a finite minimum fragment length $\lambda>0$, which allows one to model accurately the fragmentation of the whole rod within the Kirchhoff theory while avoiding the many difficulties associated with the small scale behavior around a fracture. Below, we present measurements and scaling arguments that show how $\lambda$ depends on the end-to-end bending speed.  To describe the near-adiabatic twist experiments, we adopt the empirical value $\lambda_0\approx 3.0\,$cm  measured at speed $\sim 3\,$mm/s  and zero twist.

\begin{figure*}[t!]
\includegraphics[width=2\columnwidth]{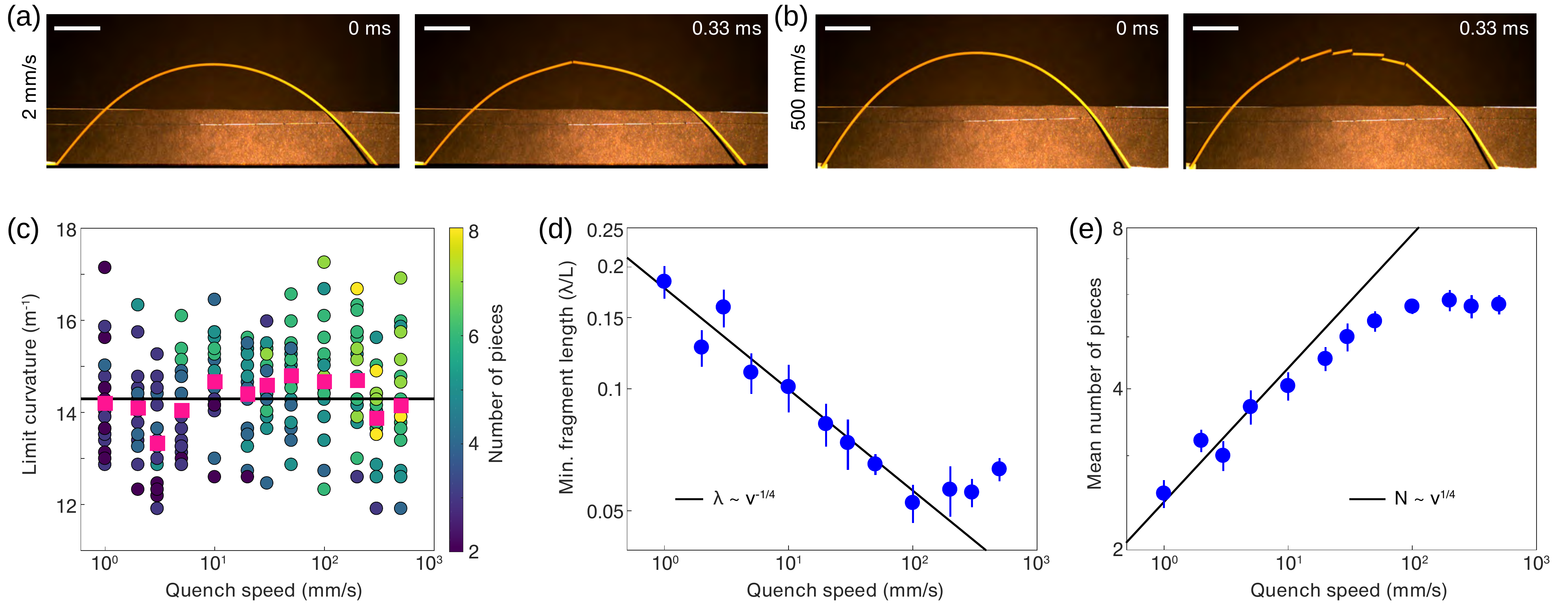}
\caption{\textbf{Dynamically quenched fracture in brittle elastic rods.} 
(a)~Experiment at low quench speed $v$ showing binary fracture (Movie 3).
(b)~Experiment at high quench speed $v$ showing fracture into multiple fragments (Movie 4), even though the limit curvature before the first fracture 
is similar to that in (a).
(c)~ Distributions of the limit curvature (mean values highlighted in pink) are not significantly affected by the quench speed $v$, but the mean number fragments increases with $v$. 
(d)~The mean length of the smallest fragments follows the theoretically predicted power law scaling.
(e)~The number of fragments approaches an asymptotic power law as expected from (d). 
At the lowest quench speed ($v=1\,\text{mm/s}$) the rod breaks into fewer than three pieces on average. Scale bars in (a,b) 3 cm. Error bars in (d,e) show standard error.}
\label{speed}
\end{figure*}
\bigskip
\par
\textbf{Twist controlled fracture.}
The first protocol explores the role of twist in bending-induced ER fracture. Twisting modes are known to cause many counter-intuitive phenomena in ER morphology~\cite{Gerbode12,Goriely06}, including Michell's instability~\cite{2013McCormick} and supercoiling~\cite{2010Brutzer_BiophysJ}. The motivation for combining twisting and bending to achieve controlled binary fracture is based on the idea that torsional modes can contribute to the first stress-induced fracture but may dissipate sufficiently fast to prevent subsequent fractures. To test this hypothesis, we built a custom device consisting of a linear stage with two freely pivoting manual rotary stages placed on both sides (Appendix \& SI Fig.1a,b). Aluminum gripping elements were attached to each rotary stage to constrain samples close to the torsional and bending axes of rotation (Appendix).  As in Feynman's original experiment~\cite{Sykes96}, we used commercially available spaghetti as test rods. To ensure reproducibility, individual rods were cut to the same fixed length $L=24\,$cm, and experiments were performed in a narrow temperature and humidity range (Appendix). The rods' ends were coated with epoxy to increase the frictional contact with the gripping elements, enabling us to twist samples to the point of purely torsional failure, which occurred at $\sim$360 degrees for our ERs.  In each individual twist experiment, a rod was loaded  into the device, twisted to a predetermined angle, and then bent near-adiabatically (end-to-end speed $<3\,$mm/s) until fracture occurred.  Select trials were recorded with a high-speed camera at 1972 fps (Appendix).

As the first main result, our experiments demonstrate that supercritical twist angles give rise to binary fracture (Fig.~\ref{twist}). 
By contrast, for small twist angles, rods are found to fragment typically into three or more pieces  (Fig.~\ref{twist}a), in agreement with Feynman's conjecture and supporting recent experimental and theoretical results~\cite{Audoly05} for the zero-twist case. For large twist angles, however, the maximum curvature before the first fracture is substantially lowered and binary fracture becomes favored (Fig.~\ref{twist}c). Although sample inhomogeneities lead to a distribution of fragment numbers at the same twist angles, the average number of fragments exhibits a robust trend towards binary fracture for twist angles larger than $\sim 250$ degrees (Fig.~\ref{twist}e,f). In particular, the experimental data follows a von Mises-type ellipsoidal curve when plotted in the plane spanned by the limit curvature and twist angle (Fig.~\ref{twist}e,f). We next rationalize these observations by performing mode analysis using the nonlinear elasticity model.

\par
We consider the dynamics after the first fracture, starting from the fact that twist enables the rod to store its energy in more than one mode. We assume the first fracture occurs at $t=0$ at the midpoint of the rod, when the curvature exceeds the critical value $\kappa_c$ determined by Eq.~\eqref{twistyield}. Our experiments and simulations show that at large twists, the rod breaks with low curvature (Fig.~\ref{twist}c-f). Focusing on this limit, we may assume that the rod is approximately planar, and that the bending is small. Under these assumptions, the twist density and bending modes uncouple (SI), and the dynamical equation for $\theta$ reduces to a damped wave equation $(\mu/\rho)\theta_{ss} = \theta_{tt} + 2b\theta_t$ for $s\in[0,L/2]$. Similarly, the small bending assumption allows us to describe the bending dynamics via the Euler-Bernoulli beam equation, $EIy_{ssss} + \rho A y_{tt} =0$, where the centerline is now given by $y(s,t)$.  Scaling arguments simplify the analysis of these equations. The speed of the twist waves is determined by shear modulus and density, $c_{\theta}= \sqrt{\mu/\rho}$. Enforcing the free-end boundary condition, $\theta'(L/2)=0$, for the undamped twist equation yields a solution with a region of zero twist stress ($\theta'=0$) growing at speed $c_\theta$ from the $s=L/2$ endpoint (SI). With non-zero damping, this picture is valid for propagation over small distances. In particular, the time taken for the zero twist stress front to travel distance $\lambda_0$, the minimum fragment length, is $T^0_\theta = \lambda_0/c_\theta$. Over longer lengthscales $\ell>\lambda_0$, the damping term becomes important. The zero twist front travels distance $\ell$ in time $T^{\ell}_\theta = b\ell^2/c_\theta^2$. Since twist modes are approximately critically damped (SI), with $b\approx \pi c_\theta/L$, we find $T^\ell_\theta = \pi\ell^2/c_\theta L$, implying that twist modes dissipate after a time $T^{\text{diss}}_\theta=L/c_\theta$. We analyse the speed of bending modes in a similar way. Over short timescales, we consider a wavepacket of bending waves. The speed of a bending wavepacket peaked at wavenumber $k$ is given by $c_b = 2k\sqrt{EI/\rho A}$. The smallest relevant length scale is the minimum fragment length $\lambda_0$, yielding the maximum allowed $k_0=2\pi/\lambda_0$. Thus the time taken for the bending wavepacket to travel distance $\lambda_0$ is $T^0_b\approx \lambda_0\left[(4\pi/\lambda_0)\sqrt{EI/\rho A}\right]^{-1} = (\lambda_0^2/2\pi r)\sqrt{\rho/E}$. Similarly, the time taken for the bending wave with wavenumber $k_0$ to travel a longer distance $\ell$ is given by $T^\ell_{k_0} = (\lambda_0\ell/2\pi r)\sqrt{\rho/E}$. However, if the time taken for the location of maximum bending stress to travel distance $\ell$ is $T_b^\ell$, then owing to dispersive effects, $T_b^\ell > T^\ell_{k_0}$. Using the estimate for $T_b^0$ and the lower bound for $T_b^\ell$, we can compare the time scales on which twist and bending operate: 
\begin{subequations}
\begin{eqnarray}
T^0_b = \frac{\lambda_0}{2\pi r}\left[\frac{1}{2(1+\nu)}\right]^{1/2} T^0_\theta 
\end{eqnarray}
and
\begin{eqnarray}
T^\ell_b > \frac{L\lambda_0}{2\pi^2\ell r}\left[\frac{1}{2(1+\nu)}\right]^{1/2} T^\ell_\theta
\end{eqnarray}
\end{subequations} 
\begin{figure*}[t!]
\includegraphics[width=2\columnwidth]{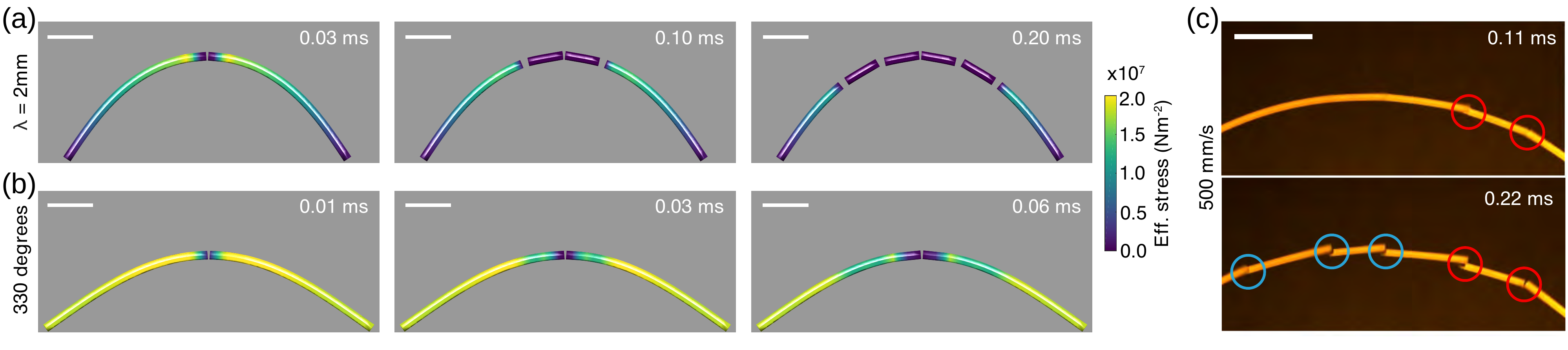}
\caption{\textbf{Simulations explain the role of twist in the fracture cascade.} 
(a)~Bending waves originating at the point of first fracture create additional fractures separated by at least the minimum fragment length~$\lambda$. In simulations, fragments are frozen after fracture and do not evolve further. 
(b)~Twist waves propagate much faster than bending waves while simultaneously lowering the critical curvature for fracture. At supercritical twist, the resulting bending waves are not strong enough to cause subsequent fractures at a spacing allowed by typical values of $\lambda$ (here $\lambda = 3\,\text{cm}$). 
(c)~Fracture cascade as observed in quench experiments at zero twist. The timings of secondary fractures (circled in blue) are in agreement with the simulations in (a). Diameter of rod and gaps between fragments enhanced for visualization in (a,b). Scale bars in (a,b) $3\,$cm. Scale bar in (c) $2\,$cm.}
\label{cascade}
\end{figure*}
Using the measured value $\lambda_0\approx 3.0\,\text{cm}$, we find $T^0_b > 4 T^0_{\theta}$ and $T^\ell_b > 2 T^\ell_\theta$ for all $\ell < L/2$, indicating that twist dissipates before bending waves can trigger another fracture. In addition, we find that $T^\text{diss}_\theta \approx 2T^0_b$, which further suggests that twist plays no role in future fracture events. The above difference in propagation times is a robust result. For example, Timoshenko theory predicts even slower bending waves than Euler-Bernoulli theory~\cite{Waves_Book}, although both beam models agree very closely in our parameter regime. To complete the argument,  we observe that all the fractures occur before reflection of the bending waves at $s=0$ becomes important. Another fracture will then be triggered if and only if $\sigma(s,t)>\sigma_c$ for any $s$ satisfying the minimum fragment length criterion and $t\in [0,t_0]$, where $t_0$ is the time for the high energy bending waves to reach $s=0$. Since twist dissipates before the bending waves become relevant, we have $\max_{t\in [0,t_0]} \sigma^2 = E^2 I \max_{t\in [0,t_0]} \kappa^2$. Let $C$ be such that $\max_{t\in [0,t_0]}\kappa = C\kappa_c$. We note that even though twist dissipates quickly, the initial twist still determines the shape of the rod at $t=0$, so $C$ is a function of $\kappa_c L, Tw$ and possibly other parameters. We calculate $C$ numerically from the Kirchhoff equations for our experimental parameters, and find to one decimal place $C=1.5$ for all relevant values of $\kappa_cL$ and $Tw$ (SI). The criterion that the rod only breaks into two pieces then takes the form $E^2IC^2\kappa_c^2<\sigma_c^2$. Using (\ref{twistyield}) to eliminate $\sigma_c$, the criterion for binary fracture becomes:
\begin{eqnarray}\label{critgrad}
\kappa_c < \frac{Tw}{\sqrt{2}L(1+\nu)\sqrt{C^2-1}}.
\end{eqnarray}
The right hand side of this inequality describes a weakly $\nu$-dependent straight line in the curvature-twist plane  (Fig.~\ref{twist}e).  Ideal ERs  that undergo their first fracture at values $\kappa_c$ and $Tw$ satisfying (\ref{critgrad}) lie below this line (purple region in Fig.~\ref{twist}e) and are expected to break into exactly two pieces. This prediction agrees well with the mean number of fragments measured in our experiments~(Fig.~\ref{twist}f). We note that while binary fracture does occur outside the critical region, it is a very low probability event. This is consistent with results of Audoly and Neukirch \cite{Audoly05} who showed that non-binary fracture occurs at zero twist. By contrast, we have demonstrated that binary fracture is almost certain in the critical region.

\bigskip
\par
\textbf{Quench controlled fracture.}
Twist fracture experiments are carried out for a fixed speed $v=3\,$mm/s in the near-adiabatic regime. To systematically explore how quenching affects ER fracture, we built a second fracture device coupling a DC stepper motor to a linear stage (Appendix \& SI Fig.~1c). By adjusting the motor velocity, we can vary 
the quench speed $v$, defined as absolute relative velocity of the ends, by more than two orders of magnitude (Fig.~\ref{speed}). Our  nonadiabatic quench protocol allows the rod to bend before fracturing, in contrast to ultra-fast diabatic protocols~\cite{Gladden05} that cause fracture by exciting buckling modes in the unbent state. Previous studies have shown that the fractal nature of fragmentation~\cite{Turcotte86} and the effects of disorder~\cite{Kun16} can give rise to universal power laws. Here, we will see that nonadiabatic quenching leads to a new class of asymptotic power law relations that involve the quench parameter $v$ and can be rationalized through scaling arguments.
\par
To investigate how quenched bending dynamics affects fracture, we performed 240 fracture experiments distributed over 12 different quench speeds $v$ ranging from $1\,\text{mm/s}$ to $500\,\text{mm/s}$. As in the twist experiments, rods were of length $L=24$\,cm and temperature and humidity were controlled to minimize variance due to environmental effects (Appendix). Select trials were recorded at 9000~fps (Fig.~\ref{speed}a,b). Generally, our experiments show that an increase in the quench speed $v$ has no significant effects on the curvature prior to fracture (Fig.~\ref{speed}c), in stark contrast to the effects of twist discussed above. Changing $v$ does however affect strongly both the minimal size of the fragments (Fig.~\ref{speed}d) and the number of fragments (Fig.~\ref{speed}e,f).
\par
To understand why quench speed (at zero twist) does not affect the limit curvature, note that the critical curvature of the samples at first fracture is of the order of  $10\,\text{m}^{-1}$ across all experiments (Fig.~\ref{speed}c). This means that the potential energy density at the first fracture is $E_P \approx EI\kappa^2 \approx 10^{-1}\,\text{J/m}$. For comparison, for a hypothetical quench speed of $v=1\,\text{m/s}$, considerably higher than realized in our experiments, the kinetic energy density is \mbox{$E_K \approx \rho Av^2/2 \approx 10^{-3}\,\text{J/m} \ll E_P$}. Hence, kinetic energy is negligible in our experiments, explaining why there is no change in the curvature at which the first fracture occurs (Fig.~\ref{speed}e). 
\par
Yet, higher quench speeds  $v$ lead to higher fragment numbers (Fig.~\ref{speed}a,b). This effect can be traced back to the fact that the minimum fragment length $\lambda$ decays with $v$  (Fig.~\ref{speed}d). We rationalize this using dimensional analysis. The dynamics of the rod are overdamped, so as the rod is quenched the force on any element scales as $F\sim v$. In one dimension, force has units of energy density so we will balance $F$ against the other fundamental energy density of the system, namely the potential energy density. From the Kirkhhoff equations, the energy density of the $k$'th bending mode scales as $E_k\sim k^4$. Dimensional arguments therefore imply the scaling $k\sim v^{1/4}$. The minimal fragment length $\lambda$ imposes an energy cut-off, $k\sim \lambda^{-1}$. Thus we obtain the scaling $\lambda\sim v^{-1/4}$ as observed in our experiments (Fig.~\ref{speed}d). We note that for very high impact energies over short timescales, the kinetic energy density will become relevant again. Since $E_K \sim v^2$, this implies the scaling $k\sim v^{1/2}$. The typical fragment size $\lambda '$ will then scale as $\lambda '\sim v^{-1/2}$. This is indeed observed for diabatically fast energy injection~\cite{Gladden05}. In one-dimensional fragmentation the number of pieces $N$ and the minimal fragment length $\lambda$ are predicted \cite{Grady10} to scale as $\lambda \sim N^{-1}$, however deviations from pure power law scaling for large $N$ are expected due to the finite total length of our samples.  Combining these scaling results, we obtain the prediction $N\sim v^{1/4}$ at small $v$, in agreement with our data~(Fig.~\ref{speed}f).  In particular, this also explains why rods can undergo binary fracture when the quench velocity is very small~(Fig.~\ref{speed}a).

\section{Summary} 

We have demonstrated two distinct protocols for achieving controlled binary fracture in brittle elastic rods. 
By generalizing classical fracture arguments~\cite{Audoly05} to account for twisting and quenching, we were able to rationalize the experimentally observed fragmentation patterns (Fig.~\ref{cascade}). Due to their generic nature, the above theoretical considerations can be expected to apply
to torsional and kinetic fracture processes in a wide range of one-dimensional structures, from construction beams~\cite{Beams_Book}  to the intracellular cytoskeleton~\cite{Tang-Schomer10,Odde99}. Indeed the appearance of a power law response to quenching is a common feature of many results in condensed matter physics~\cite{Young2016,Retzker2010}. While our results demonstrate two concrete loopholes for violating Feynman's conjecture, they also 
suggest several directions for future research. New theory beyond the Kirchhoff model is needed to clarify the microscopic 
origin of the minimum fragment length and to explain the nonplanar geometry of the fracture interfaces. From a practical perspective, 
it will be interesting to explore whether, and how,  twist can be utilized to control the fracture behavior of two- and three-dimensional materials. 
\par
\textbf{Acknowledgments}
We thank Dr. Jim Bales (MIT) for providing the high speed cameras.
This work was supported by an Alfred P. Sloan Research Fellowship (J.D.) and a Complex Systems Scholar Award from the James S. McDonnell Foundation (J.D.).

\appendix
\section{Experiments}
All experiments used Barilla no.~3 raw spaghetti of length $L=24\,$cm. High air humidity and large temperature fluctuations 
can affect bending  and fracture behavior of the samples. Throughout our experiments, air humidity was kept low in the range 21\%-34\%. 
Temperature during the twist experiments was kept constant at $22.5\pm 1.5\,$C and during the kinetic quench 
experiments at $25.5\pm 0.5\,$C. The rod diameter was measured using calipers for 5 samples to give $2r = 1.4 \pm 0.1\,$mm. The density was obtained by weighing 10 samples cut to $24\,$cm. Treating the samples as cylinders of radius $r$ we obtained $\rho = 1.5 \pm 0.1\, \text{g/cm}^3$. The Young's modulus, $E$, was measured by applying a slowly increasing longitudinal compression force to samples positioned upright upon a scale. The sample length $\ell$ and the mass $m$ shown on the scales at the point of buckling were recorded. We repeated this for 20 samples of varying lengths, and in each case calculated $E$ from the Euler buckling criteria, $mg=\pi^2EI/\ell^2$, to find $E=3.8\pm 0.3\,$GPa. The shear modulus was measured by attaching a mass of known moment of inertia, $I_0$, to samples of varying length to create a torsion pendulum. The angular frequency, $\omega^2=\mu J/I_0 L$, was obtained for 5 samples, by filming the pendulum with an Edgertronic SC2 at 1972 fps. The Poisson's ratio, $\nu=(E/2\mu) -1$ was calculated from the mean values of $E$ and $\mu$. Using standard error of these mean values to quantify uncertainty gives $\nu = 0.3\pm 0.1$. The twist damping parameter was obtained from the decay rate of the torsion pendulum (SI).
\par
Twist experiments were performed using a custom-built device comprising of a manual linear stage with two freely pivoting manual rotary stages placed on both sides. Aluminum gripping elements were attached to each rotary stage to constrain samples close to the torsional and bending axes of rotation (SI). We completed 73 trials at various twist angles up to 360 degrees, corresponding to the approximate pure torsion yield stress of the samples. To ensure proper reproducible twisting of the samples within the desired range, the ends of each rod were coated with Devcon® 5 minute epoxy gel. The epoxy increased friction between each sample and the gripping element in our testing device, enabling us to twist samples to the point of torsional failure. Each sample was then loaded into the device, twisted to the chosen angle, and bent until fracture occurred. The ends were moved together slowly ($< 3\,$mm/s) to ensure a quasi-static regime. The end-to-end distance at the onset of fracture was recorded for each trial. Select trials were recorded with an Edgertronic SC2 at 1972 fps.  
\par
Kinetic quench experiments consisted of 20 trials each for 12 speeds ranging from $1\,\text{mm/s}$ to $500\,\text{mm/s}$, using a custom single-axis linear stage controlled by a NEMA 17 Bipolar DC Stepper Motor (SI). The device moved the ends of each sample towards each other at a fixed speed $v$ while allowing them to pivot freely as bending occurred. The end-to-end distance at the onset of fracture was obtained by recording each trial with a Photron FASTCAM Mini AX200 at 9000 fps and examining the playback. Select trials were stored permanently.

\section{Experiments}
Numerical results in Fig.~\ref{twist}b,d were obtained by simulating the Kirchhoff equations~\eqref{Kirch} 
using a discrete differential geometry algorithm~\cite{Bergou08,Bergou10}. Each rod was discretized into 50 elements and one time-step of simulation time corresponded to $1\mu$s of real time. Time-stepping was performed with a Verlet scheme. Fracture was simulated by disconnecting the rod in 1 time-step wherever the above fracture criterion was satisfied.  The radius of the rod has been enhanced in simulation images (Fig.~\ref{twist}b,d and Movies 1,2) for visualization purposes. The curvatures in Fig.~\ref{twist}e,f and Fig.~\ref{speed}c were obtained numerically from the observed end-to-end distance by initializing a rod with the appropriate boundary conditions and twist, and allowing it to relax to its lowest energy state via gradient descent. In the case of zero twist, there is a closed form relationship between end-to-end distance and maximum curvature, which we used to validate the code (SI). The Matlab code is available on request. 



\end{document}